\documentclass[12pt]{article}
\usepackage{graphicx}

\newcommand\pubnumber{NOBUGS2002/019}
\newcommand\pubdate{\today}

\def\frm2{ZWE FRM-II\\
Technische Universit\"at M\"unchen\\
Lichtenbergstr. 1\\
D-85748 Garching, F.R.G.\\
jens.krueger@frm2.tu-muenchen.de\\
}

\def\Title#1{\begin{center} {\Large #1 } \end{center}}
\def\Author#1{\begin{center}{ \sc #1} \end{center}}
\def\Address#1{\begin{center}{ #1} \end{center}}

\newcommand\pubblock{\rightline{\begin{tabular}{l} \pubnumber\\
         \pubdate  \end{tabular}}}
\newenvironment{Abstract}{\begin{quotation}  }{\end{quotation}}
\newenvironment{Presented}{\begin{quotation} \begin{center} 
             PRESENTED AT\end{center}\bigskip 
      \begin{center}\begin{large}}{\end{large}\end{center} \end{quotation}}

\def\beq{\begin{equation}}
\def\eeq#1{\label{#1}\end{equation}}
\def\eeqn{\end{equation}}

\def\beqa{\begin{eqnarray}}
\def\eeqa#1{\label{#1}\end{eqnarray}}
\def\eeqan{\end{eqnarray}}

\let\bar=\overbar

\def\Dslash{\not{\hbox{\kern-4pt $D$}}}
\def\dslash{\not{\hbox{\kern-2pt $\del$}}}

\def\msb{{\bar{\ssstyle M \kern -1pt S}}}

\begin{document}
\begin{titlepage}
\pubblock
\vfill
\Title{Instrument Control with TACO at the FRM-II}
\vfill
\Author{Jens Kr\"uger,  J\"urgen Neuhaus}
\Address{\frm2}
\vfill

\begin{Abstract}
With the start up of the new German Neutron Source FRM-II in Garching,
a common instrument control system has to be installed at the
different neutron scattering instruments. With respect to the large
variety of instruments, a modular system was required. In spite of
starting from scratch, a well developed and tested system was
needed, in order to meet the restricted man power and short time
scale to realise the project. We decided to adopt the TACO system from
the ESRF in Grenoble. The main arguments were the network transparent
communication of modules, the proven reliability and the support by
the developer team at the ESRF.
\end{Abstract}
\vfill
\begin{Presented}
NOBUGS 2002\\
Gaithersburg, U.S.A,  November 04--06, 2002
\end{Presented}
\vfill
\end{titlepage}

\def\thefootnote{\fnsymbol{footnote}}
\setcounter{footnote}{0}

\section{Introduction}

Our decision for a common instrument control system at the FRM-II was based on the analysis of the required
hardware we have to support. Modern electronic equipment at large scale user facilities can be devided into two
categories: slow control and high speed data acquisition. A large part of the instrument for neutron 
scattering experiments belongs to the first category. These are all motor controls, switches, or sample environment
equipments which change their state only slowly, that means in time scales of seconds or even longer. 

\section{Hardware}

These slow controls are generally connected by field-busses, i.e. serial lines or Profibus DP in our case. The main
advantage of using the industry standard Profibus is to connect a large variety of hardware by the same interface.
Furthermore one has to deal with a complex protocol if the own hardware is to be developed. For this
work we rely on a collaboration with the electronic department (ZEL) of the Forschungszentrum J\"ulich, Germany 
(see contribution by H. Kleines \cite{Kleines}).
For a large number of commercially available control electronics (vacuum systems, power supplies, high voltage supplies, ...),
the only available interface is the RS232 connection.
If the number of devices gets too large we group systems together by a RS485 field-bus using a simple modbus protocol
to address the single device.

For high speed data acquisition like position sensitive linear or area
detectors or even single counting boards we use faster computer busses
like PCI or VME, depending on the complexity of the electronics. For
our own developments we prefer to use so called M-Modules (M-Module
Mezzanine (ANSI/VISTA 12-1996)\cite{MModule}) boards which can be
plugged in commercially available carrier boards for PCI, cCPI, or VME
systems.

\section{Software requirements}

The instrument control has to fulfill two major tasks:
\begin{itemize}
\item communication with the electronic hardware
\item interface to the user
\end{itemize}

As there are hardly any interferences between these tasks, we aimed at
a modular system where different components of the instrument can be
put together without restrictions on the number or location of the
components.

\subsection{TACO - an object oriented control system}

\begin{figure}[htc]
\begin{center}
\includegraphics[height=8cm]{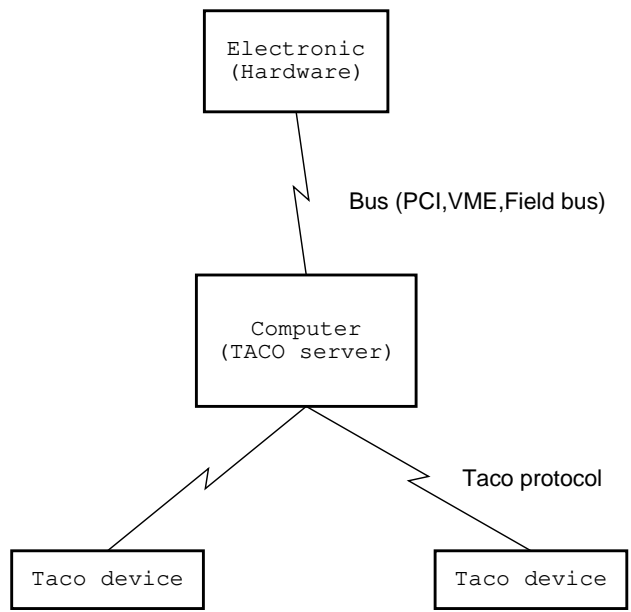}
\end{center}
\caption{Schematic layout of the TACO system communication.}
\label{taco-hardware}
\end{figure}

Such a system has been realised at the ESRF\cite{esrf} for the beam
line control and is now used as well for their instrument control.
This so called TACO system \cite{TACO} presents the electronic
components as network transparent devices. The TACO system itself
provides the organization and communication to these devices (see
figure \ref{taco-hardware}).

The TACO servers which export devices, on the one hand ensure the
communication to the hardware (via field-busses, or PCI/VME
communication) and on the other hand provide instrument parameters in
physical units, e.g. sample temperature in Kelvin, motor movements in
centimeters, or axis movements in degrees, and so on). This simplifies
the access to the hardware for the user, without knowledge of the hardware 
internals beeing required.

One of the striking features of TACO is that the communication between
the different modules is not restricted to a specific operating system
on the computers as TACO uses the SunRPC communication implemented on
the different UNIX flavors as well as on Windows systems. However,
for simplicity we restricted the development at the FRM-II to Linux so
far. This decision implies that for a large number of commercially
available hardware, Linux drivers had to be developed by ourselves,
which finally leads to a natural selection of the used hardware.

\subsection{TACO extensions at the FRM-II}

\begin{figure}[htc]
\begin{center}
\includegraphics[height=8cm]{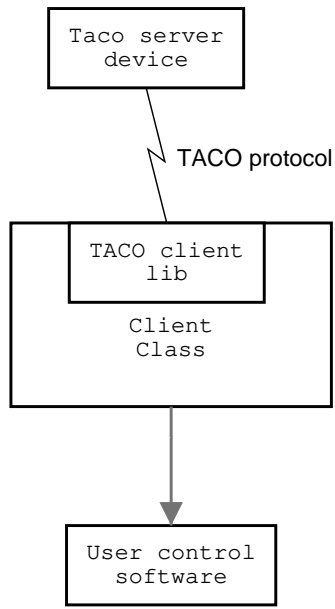}
\end{center}
\caption{Schematic layout of the TACO extention from the FRM-II.}
\label{client-class-concept}
\end{figure}

At the FRM-II TACO is not used in the form distributed by the ESRF,
but in an extended way. The TACO communication between client and
server is done by a generic protocol, which contains a unique command
number, a pointer to the input data, an information about the type of
the input parameter, and a pointer and type of the output parameter.
However, programming this communication might lead to a large number
of pitfalls, which then gives an unexpected behaviour of the control
software. We decided to implement a safer way to communicate between
client and server. The communication between them is hidden by so
called client classes, which are created by the developer of the
server (see figure \ref{client-class-concept}). These classes are counterparts to
the server exported devices. They support all commands provided by devices
and give the user a safe communication way.  Additionally it tests
whether the client has connected to a corresponding device.

Another extension used at the FRM-II is the concept of high level or
logical TACO servers. These servers have no direct connection to the
hardware. They use the low level taco server as clients, create a
logical unit and provide corresponding TACO devices. By this we get a
lot of complex devices, such a axis (a motor which is mechanically
connected to an encoder device), or sample environment servers which
hide the internals from the user.
 
\section{Instrument control at FRM-II}

The instrument control is realised as an Ethernet-subnet at each
instrument using standard UDP or TCP communications via SunRPC calls,
as defined in the TACO system.

The communication to the different servers by a conventional network
replaces in some way the extensive use of field-busses as in the
industrial automation. Depending on the complexity of the instrument,
electronics is connected to a range of different computers.

\begin{figure}[htc]
\begin{center}
\includegraphics[height=8cm]{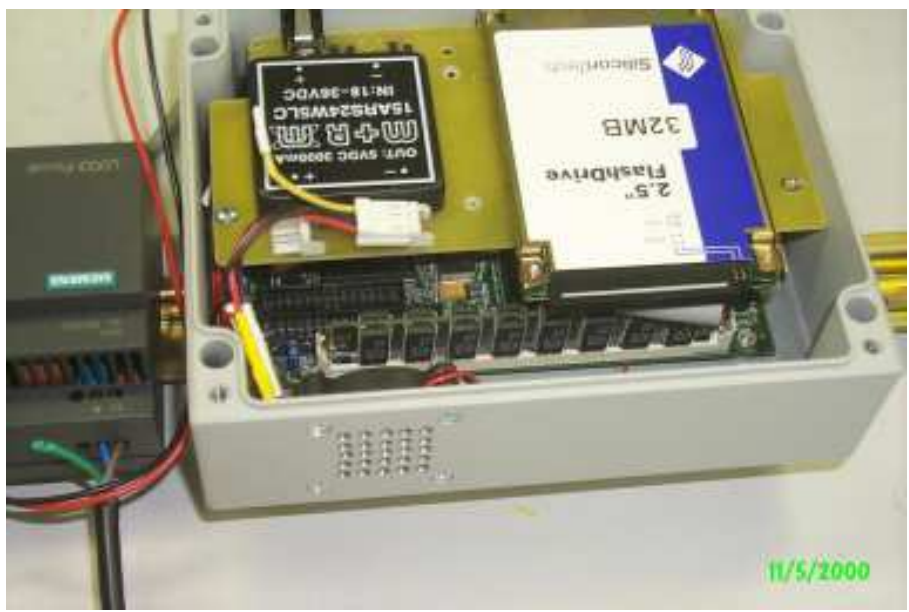}
\end{center}
\caption{View into an open so-called TACO box}
\label{taco-box}
\end{figure}

For electronical and mechanical stability we prefer the compact PCI
standard as hardware platform. For special purposes like a high
temperature furnace we group all the connections to a tiny single
board computer with a flash disk (see figure \ref{taco-box}), put it in a nice cave and call it
according to the ESRF a TACO box. For this we developed a tiny
linux distribution, based on a SuSE 7.2 distribution.

\subsection{User interface}

For the user interface we developed a front-end written in python
\cite{python}, a common scripting language, called NICOS \cite{NICOS}
(see contribution by T. Unruh \cite{tunruh}). It provides full remote
control of the instrument in a client/server architecture.

\subsection{Data displaying}

A powerful instrument control should also provide an online data
visualization. For this we started to develop a new frame work with a
modular architecture, to adopt it in a simple way to the different
needs of the instruments. We called it openDaVE (open Data
Visualization Environment) \cite{opendave} (see contribution openDaVE
by J .Kr\"uger \cite{jkrueger}).

\subsection{Data format}

Throughout the instruments we want to establish a common data format,
i.e. NeXus, which is the major data format for openDaVE(see contribution by J. Beckmann \cite{beckmann}).







 
\end{document}